\documentclass[aps, reprint, prl, twocolumn, superscriptaddress, longbibliography]{revtex4-1}
\usepackage{times}
\usepackage{amsmath}
\usepackage{amssymb}
\usepackage{graphicx}
\usepackage{float}
\usepackage{braket}
\usepackage{xcolor}
\usepackage[utf8x]{inputenc}
\usepackage[colorlinks=true, urlcolor=blue, linkcolor=blue, citecolor=blue]{hyperref}

\newcommand{\affcua}{MIT-Harvard Center for Ultracold Atoms, Research Laboratory of Electronics, and Department of Physics, Massachusetts Institute of Technology, Cambridge, Massachusetts 02139, USA}
\newcommand{\affens}{D\'{e}partement de Physique, Ecole Normale Sup\'{e}rieure / PSL Research University, CNRS, 24 rue Lhomond, 75005 Paris, France}

\begin{document}

\title{Universal Sound Diffusion in a Strongly Interacting Fermi Gas}

\author{Parth B. Patel}
\affiliation{\affcua}
\author{Zhenjie Yan}
\affiliation{\affcua}
\author{Biswaroop Mukherjee}
\affiliation{\affcua}
\author{Richard J. Fletcher}
\affiliation{\affcua}
\author{Julian Struck}
\affiliation{\affcua}
\affiliation{\affens}
\author{Martin W. Zwierlein}
\affiliation{\affcua}

\date{\today}

\begin{abstract}
Transport of strongly interacting fermions governs modern materials -- from the high-$T_c$ cuprates to bilayer graphene --, but also nuclear fission, the merging of neutron stars and the expansion of the early universe.
Here we observe a universal quantum limit of diffusivity in a homogeneous, strongly interacting Fermi gas of atoms by studying sound propagation and its attenuation via the coupled transport of momentum and heat.
In the normal state, the sound diffusivity ${D}$ monotonically decreases upon lowering the temperature $T$, in contrast to the diverging behavior of weakly interacting Fermi liquids.
As the superfluid transition temperature is crossed, ${D}$ attains a universal value set by the ratio of Planck's constant ${h}$ and the particle mass ${m}$.
This finding of quantum limited sound diffusivity informs theories of fermion transport, with relevance for hydrodynamic flow of electrons, neutrons and quarks.
\end{abstract} 

\maketitle 


\begin{figure*} 
\includegraphics[width=2\columnwidth]{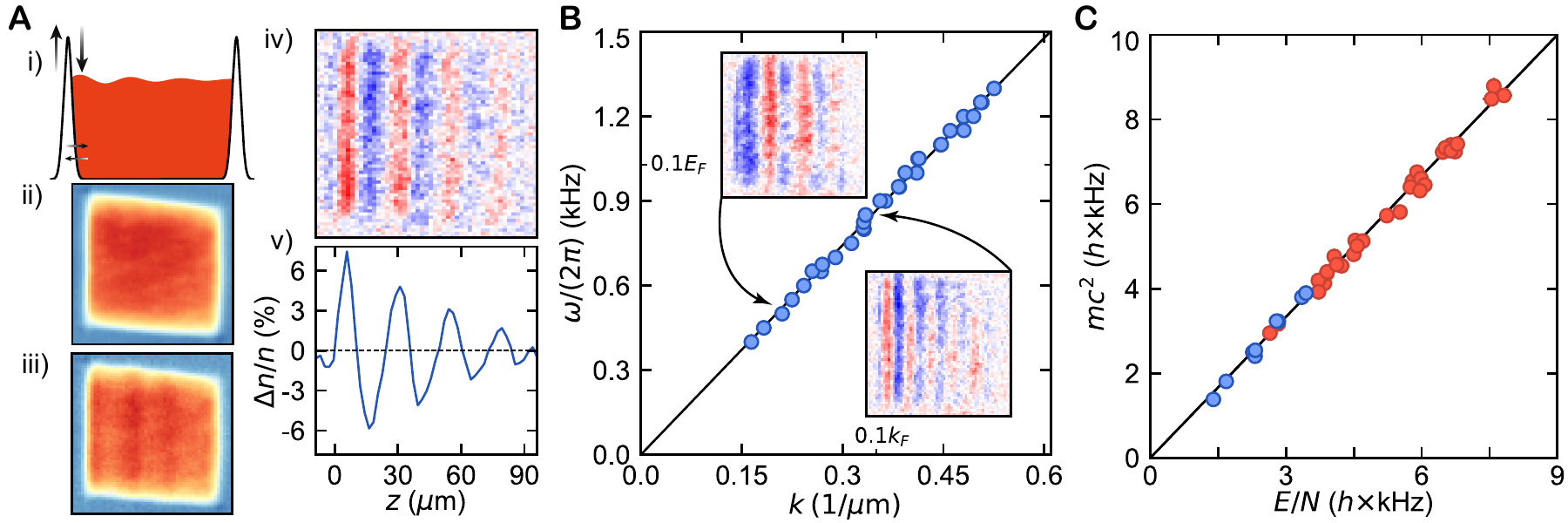}
\caption{\textbf{Sound waves in a homogeneous unitary Fermi gas.}
(\textbf{A}) Sound is excited by modulating the intensity of one of the laser walls (i) and the resulting density wave is observed via an in situ absorption image, shown for both an unperturbed (ii) and modulated (iii) sample. Here the modulation frequency is $2\pi\times 600~$Hz. Taking their difference (iv) and integrating along the homogeneous radial trap axis reveals (v) a perturbation in the fractional density difference $\Delta n/n$, propagating along the axial direction $z$ and exhibiting a well-defined wavenumber $k$ corresponding to the applied modulation frequency $\omega$.
(\textbf{B}) Dispersion of sound $\omega(k)$. The fitted slope (black line) provides the speed of sound. The insets display sound waves observed at $\omega=2\pi\times 500\,\rm Hz$ and $2\pi \times 850$ Hz. Errors in the measured $k$ are smaller than the point size. 
(\textbf{C}) Measurement of the universal relation between the measured speed of sound and the energy-per-particle $E/N$ (see text). The black solid line shows the predicted linear dependence for any non-relativistic scale invariant system in 3D; $mc^2 = \frac{10}{9} E/N$. Data are shown for both the normal (red) and the superfluid (blue) phase. 
 \label{Fig1}}
\end{figure*}

Transport in fermionic quantum matter lies at the heart of phenomena as varied as superconductivity in cuprates~\cite{PatrickLee2006:HighTcReview} and bi-layer graphene~\cite{CaoPablo2018:MagicAngleGraphene}, inspirals of neutron star binaries~\cite{Alford2018:ViscosityNeutronStar} and perfect fluidity of the early universe~\cite{SchaferThomas2012:QuantumFluids}.
For hydrodynamic flow, transport is governed by diffusion, via which spatial variations in globally conserved quantities, such as momentum, energy, charge, or spin, decay at a rate set by the corresponding diffusivity. 
A ubiquitous example is the attenuation of sound in fluids, where the modulation in current density and temperature causes diffusion of momentum and heat, leading to attenuation of sound at a rate set by the sound diffusivity $D$.
The magnitude and temperature dependence of sound diffusivity reveal many of the characteristic features of the underlying substance.

Kinetic theory yields an estimate of ${D\simeq v l}$, where $v$ is the average velocity of the particles and $l$ their mean-free path, which can vary over many orders of magnitude between substances. 
However, for strongly interacting quantum liquids and gases, a certain universality of diffusion coefficients may be expected. Here, the mean-free path becomes on the order of the interparticle spacing, and the velocity takes on the Heisenberg-limited value ${v\sim \hbar/m l}$ leading to a limiting value of ${D\sim \hbar/m}$, independent of the details of the microscopic interactions~\cite{SommerZwierlein2011:SpinTransport}.
Indeed, such limiting values are observed for the spin diffusivity in a unitary Fermi gas~\cite{SommerZwierlein2011:SpinTransport}, as well as the momentum diffusivity (the kinematic viscosity) in both the quark-gluon plasma of the early Universe and the unitary Fermi gas~\cite{SchaferThomas2012:QuantumFluids}.
Remarkably, the quantum liquids of bosonic $^4$He and fermionic $^3$He display similar sound diffusivities of $D\sim \hbar/m$ around $4$~$K$ ~\cite{KettersonBennemannHeliumBook,VollhardtWolfleHe3Book}. 
However, upon lowering the temperature into the deeply degenerate regime, these two quantum liquids display strikingly different behaviours in the damping of sound.
Down to about one Kelvin, the sound attenuation in $^4$He does not vary strongly with temperature, decreasing only by a factor of two across the superfluid transition, with a minimum of $D\simeq 0.5 \hbar/m$~\cite{KettersonBennemannHeliumBook,LvovSreenivasan2014:4HeViscosityCirculation}. 
On the other hand, $^3$He features a diverging diffusivity $\left( \propto 1/T^2 \right)$, characteristic of a Fermi liquid, growing to ${\sim \! 50,\!000 \, \hbar/m}$ around $2$~mK, followed by a steep drop at the superfluid transition and settling to a value of ${\sim\! 5,\!000\,\hbar/m}$~\cite{VollhardtWolfleHe3Book}.
\emph{A~priori}, it is unclear whether the temperature dependence of sound attenuation in a strongly interacting, fermionic gas -- of atoms, electrons or neutrons -- should resemble at all that of a quantum liquid; and if so, whether it corresponds more closely to the strongly interacting, but bosonic, liquid $^4$He or to the fermionic, but weakly interacting, liquid $^3$He.

Ultracold atomic Fermi gases at unitarity are a prototypical strongly interacting quantum fluid for transport experiments~\cite{Ketterle2008:MakingProbingFermiGas,Bloch2008:ManyBodyPhysics,Giorgini2008:FermiGasTheory,ZwergerBook2012,Zwierlein2014:NovelSuperfluids}. 
Featuring a mean free path as short as one interparticle spacing, these systems display the most robust form of fermionic superfluidity and near-perfect hydrodynamic flow even in the normal state~\cite{Thomas2002:HydroAnisotropicExpansion,Altmeyer:2007,GrimmZwierleinStringari2013:CollectiveModes}.
The presence of scale invariance leads to universal physics~\cite{Ho2004:UniversalThermo,Giorgini2008:FermiGasTheory,ZwergerBook2012,KuZwierlein2012:EoS,Zwierlein2014:NovelSuperfluids,Randeria2014:BECBCS} and transport properties~\cite{CaoJohnThomas2011:Viscosity,SommerZwierlein2011:SpinTransport,EnssZwerger2011:Viscosity,JosephThomas2015:ViscosityNearTc}, offering a direct connection to a host of strongly interacting Fermi systems across all energy and length scales from nuclear matter to neutron stars.
For the unitary Fermi gas, scale invariance implies that sound diffusivity must remain the same upon changing all length scales by the same factor. The diffusivity is thus $\hbar/m$ times a universal function of $T/T_\textrm{F}$, the temperature $T$ normalized by the Fermi temperature $T_\textrm{F}$ that only depends on the particle density $n$~\cite{BrabySchafer2010:ThermalCondSoundAtten}. 
At non-degenerate temperatures $T \gg T_\textrm{F}$, we expect a unitary Boltzmann gas, where the thermal wavelength ${\lambda=\sqrt{2 \pi \hbar^2 / (m k_\textrm{B} T)}}$ sets both the mean free path and the typical velocity of excitations, ${l\sim 1/(n\lambda^2)}$ and ${v\sim \hbar/(m\lambda)}$, implying ${D \sim (\hbar/m) (T/T_\textrm{F})^{3/2}}$.
In the quantum critical regime of the unitary gas~\cite{NikolicSachdev2007:QmLiquidsCritical,Enss2012:QmCritTransportUFG}, at $T \sim T_\textrm{F}$, the interaction and thermal energies are comparable and even the nature of the equilibrium state is a subject of debate~\cite{NascimbneSalomon2011:FermiLiquidUFG,RothsteinShrivastava2019:NoFermiLiquidUFG}. 
At low temperatures $T\ll T_\textrm{F}$, it remains unknown whether the sound diffusivity diverges as $1/T^2$~\cite{TransportPhenomenaBook:JensenSmith}, as in the Fermi liquid   $^3$He~\cite{BlackThompson1971:FermiLiquid3He,ArchieRichardson1981:Viscosity3HeNormal}, and whether any sudden drop in the sound diffusion occurs upon entering the superfluid regime.
Predictions for the kinematic viscosity vary from zero~\cite{GuoLevin2011:PerfectFluidBadMetals}, as suggested by experiments on expanding inhomogeneous gases~\cite{CaoJohnThomas2011:Viscosity,JosephThomas2015:ViscosityNearTc}, to infinity if phonon damping dominates~\cite{KurkjianCastin2017:PhononInteractions,EnssZwerger2011:Viscosity}.

Transport experiments on Fermi gases have thus far employed harmonic traps~\cite{Zwierlein2014:NovelSuperfluids} or terminal configurations~\cite{KrinnerEsslingerBrantut2017:TransportReview,ValtolinaRoati2015:JosephsonFermiGas}, and have been used to probe collective oscillations~\cite{BartensteinGrimm2004:CollectiveOsc,KinastThomas2004:EvidenceSFCollectiveOsc,HoinkaVale2017:GoldstonePairBreakingExcitationInUFG}, spin transport~\cite{SommerZwierlein2011:SpinTransport,KoschorreckKohl2013:SpinDynamics2DFermi,LuciukThywissen2017:SpinTransport}, viscosity~\cite{CaoJohnThomas2011:Viscosity}, conductivity~\cite{KrinnerEsslingerBrantut2017:TransportReview} and Josephson oscillations~\cite{ValtolinaRoati2015:JosephsonFermiGas}. 
However, obtaining transport coefficients of homogeneous matter from inhomogeneous samples in atom traps requires sophisticated analysis and assumptions on the spatial flow profile~\cite{SommerZwierlein2011:SpinTransport,CaoJohnThomas2011:Viscosity}.
With the recent advent of optical box traps~\cite{GauntZoran2013:HomogBEC,MukherjeeZwierlein2017:HomogFermi,HueckMoritz2018:HomogFermi2D,BairdThomas2019:HydroLinResponseUFG}, it is now possible to directly probe the transport properties of homogeneous quantum gases~\cite{NavonZoran2016:TurbulentCascade,VilleDalibard2018:2DSoundBEC,GarrattZoran2019:BoxExcitationsBEC,BairdThomas2019:HydroLinResponseUFG}.
The gas is then in the same state throughout and transport properties are identical across the system.

Measurements of transport properties involve the response of a system to an external drive. 
In linear response, an applied potential $V$ couples to perturbations in the fluid density ${\delta  n= \chi V}$ via the density response function $\chi$.
Sound corresponds to a resonant response, that is a pole in $\chi$ at a frequency ${\omega=ck}$, set by the speed of sound $c$ and wavenumber $k$, in the vicinity of which $\chi(\omega,k) \sim {1/\left( \omega^2 - c^2 k^2 + i \Gamma \omega \right)}$~\cite{Hohenberg1965:SFHeTheory,Hohenberg1973:DensityCorrFunc}.
Here, $\Gamma$ is the damping rate of sound, given by ${\Gamma = D k^2}$~\cite{LL:FluidMechanics} for hydrodynamic systems. 
Measurements of $\chi$ and $\Gamma$ thus directly provide the sound diffusivity.
Experiments involving liquid helium have used a number of techniques to measure $\chi$, from free decay of resonant modes in a cylindrical resonator~\cite{KettersonBennemannHeliumBook,VollhardtWolfleHe3Book} to Brillouin scattering off of sound waves~\cite{TarvinGreytak1977:StructureFactor}.

\begin{figure*}
\includegraphics[width=1.8\columnwidth]{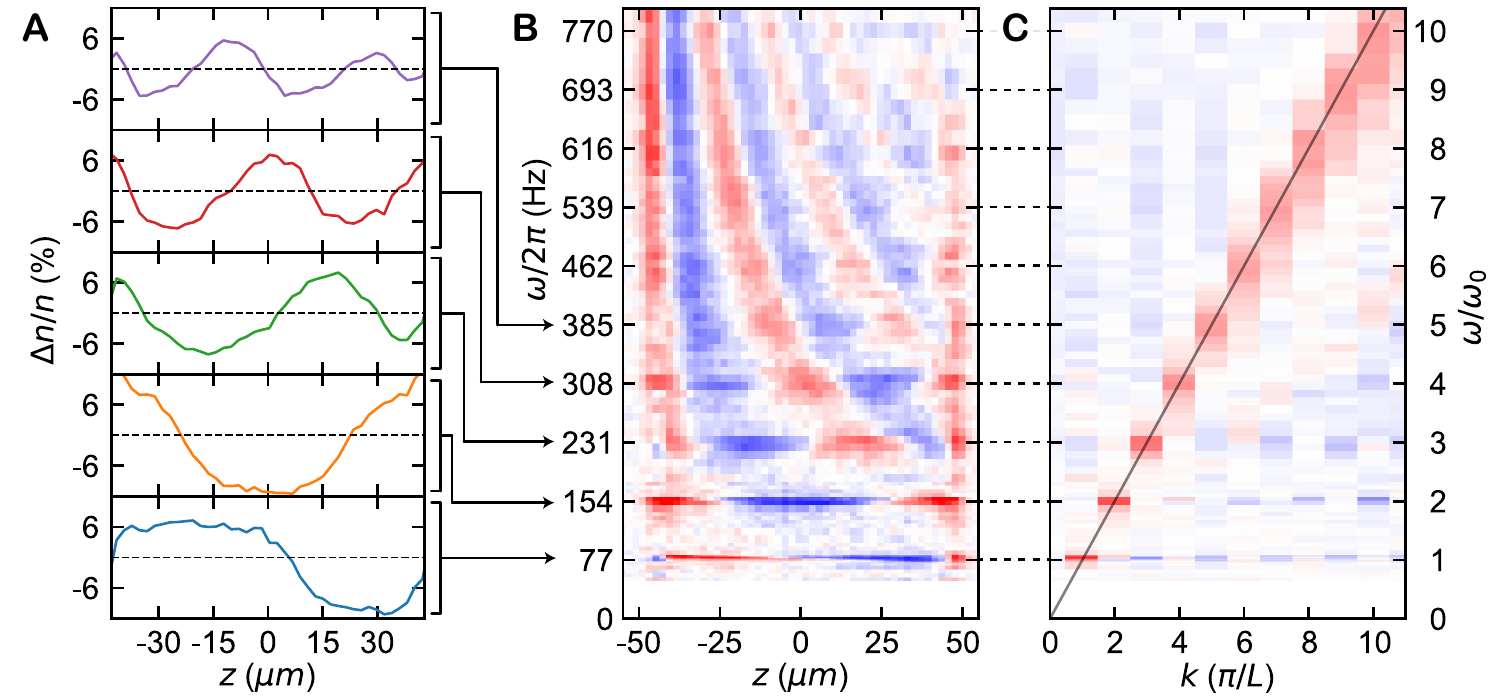}
\caption{\textbf{Normal modes of the cylindrical box trap.} 
The steady state density response of the gas is obtained by modulating the container walls at frequency $\omega$ for 30 cycles of the drive. Standing waves of sound corresponding to the normal modes in the box are observed at frequencies $\omega_{j} =  j \pi c / L \approx 2 \pi j\times 77$ Hz (where $j\in\mathbb{Z}$), the first five of which are shown in (\textbf{A}). The full sonogram is shown in (\textbf{B}). Here, each row of pixels corresponds to a particular realization of the experiment at a given frequency. The spatial Fourier transform directly yields the density response function $\textrm{Im}[\chi(k,\omega)]$ (\textbf{C}). It reveals well-defined resonance peaks exhibiting both the linear dispersion of sound and increasing widths in frequency at higher wavenumbers, corresponding to increased rates of sound attenuation.
\label{Fig2}}
\end{figure*}

In our homogeneous quantum gas, the constant background density enables an ideal realization of a density response measurement (see Fig.~1A).
We employ an equal two-state mixture $^6$Li atoms with resonant interstate interactions, confined to a cylindrical optical box potential composed of three repulsive laser beams: a hollow cylindrical beam providing the radial confinement (radius ${60~\mu}$m), and two sheets of light serving as endcaps (length $L\sim\! 100~\mu$m)~\cite{MukherjeeZwierlein2017:HomogFermi}. 
The number ${N \sim 10^6}$ of atoms per spin state yields a Fermi energy of ${E_\textrm{F} = \hbar^2 k_\textrm{F}^2/(2m) \sim h\times10~}$kHz.
To inject sound waves, we sinusoidally modulate the intensity of one endcap beam, which drives the gas at a well-defined frequency $\omega$, and a wide range of spatial wavenumbers, Fourier limited by the width ${\sim 4\mu{\rm m}}$ of the endcap potential's edge (see Supplementary Information). 
At the given driving frequency, the resonant sound response of the gas is dominated by a specific wavenumber ${k = \omega/c}$, resulting in a traveling wave of sound.
An \emph{in situ} absorption image is taken after an evolution time sufficiently short such that no reflections occur, and the resonant wavenumber $k$ is directly measured (Figs.~1A(iii-iv)).
By repeating this protocol for different drive frequencies, we obtain the dispersion relation $\omega(k)$ for wavenumbers ${k < 0.14 k_F}$. (Fig.~1B). 
It is linear within our measurement error, corresponding to a constant speed of sound ${c=\omega/k}$ as a function of wavenumber. 
We note that at wavelengths approaching the interparticle spacing, and thus at momenta $\hbar k$ approaching the Fermi momentum ${\left( k \sim k_\textrm{F} \right)}$, deviations from linear sound dispersion are expected for the unitary Fermi gas~\cite{KurkjianCastin2016:DispersionConcavity}.

The precise measurement of the speed of sound allows a sensitive test of scale invariance of the unitary Fermi gas. In general, the speed of isentropic sound propagation is directly tied to the equation of state via the hydrodynamic relation ${mc^2 = \left. ({\partial P}/{\partial n} ) \right|_S = \left. ({V^2}/{N}) ({\partial^2 E}/{\partial V^2}) \right|_S}$. Here, $E$ is the energy, $S$ is the entropy, $V$ is the volume, and ${P = \left. - ({\partial E}/{\partial V}) \right|_S}$ is the pressure of the gas. 
A remarkable property of \emph{all} non-relativistic scale invariant systems in 3D is that their total energy scales as $E \propto V^{-2/3}$; this follows from the scaling behavior $E\rightarrow E/\lambda^{2}$ under dilation of space by a factor $\lambda$.
This directly yields $m c^2 = ({10}/{9}) E/N $, independent of temperature or the phase of matter.
In Fig.~1C we show the measured speed of sound as a function of the energy per particle $E/N$, obtained from an isoenergetic expansion of the gas from the box into a harmonic trap~\cite{YanZwierlein2019:FermiLiquid}. 
For both superfluid and normal samples (blue and red, respectively), the scale invariant prediction (solid black line) captures the data well with no free parameters. This demonstrates the universality of the speed of sound and scale invariance in the unitary Fermi gas in the explored window of temperature.


The attenuation of sound is already apparent in the spatial decay of the travelling waves shown in Fig.~1.
For a precision measurement of the sound diffusivity, we now turn to the steady state response of the system to a continuous drive, which directly reveals the density response function $\chi$. 
The intensity of one of the endcap laser walls is modulated for a sufficiently long time such that the evolution has reached the steady state regime. 
After an integer number of driving cycles, the spatial Fourier transform of the density yields the out-of-phase response of the system, or $\textrm{Im}[\chi(\omega,k)]$ (see Supplementary Information). 
This quantity also gives the average power absorbed by the system for a drive at frequency $\omega$ and spatial frequency $k$, and thus directly reveals the poles of $\chi$ as resonances.
The measurements are summarized in Fig.~2.
Each row of pixels in Fig.~2B shows the fractional density modulation at a particular drive frequency after integration along the radial axis.
This `sonogram' reveals discrete normal modes, the first five of which are shown in Fig.~2A.
The spatial Fourier transform, giving the out-of-phase response function, is shown in Fig.~2C. For each normal mode in the box, it features a peak at ${\omega=ck}$.
The sound attenuation rate can be seen to increase with $k$, revealed in both a broadened frequency response as well as a reduced peak height.

\begin{figure}
\includegraphics[width=\columnwidth]{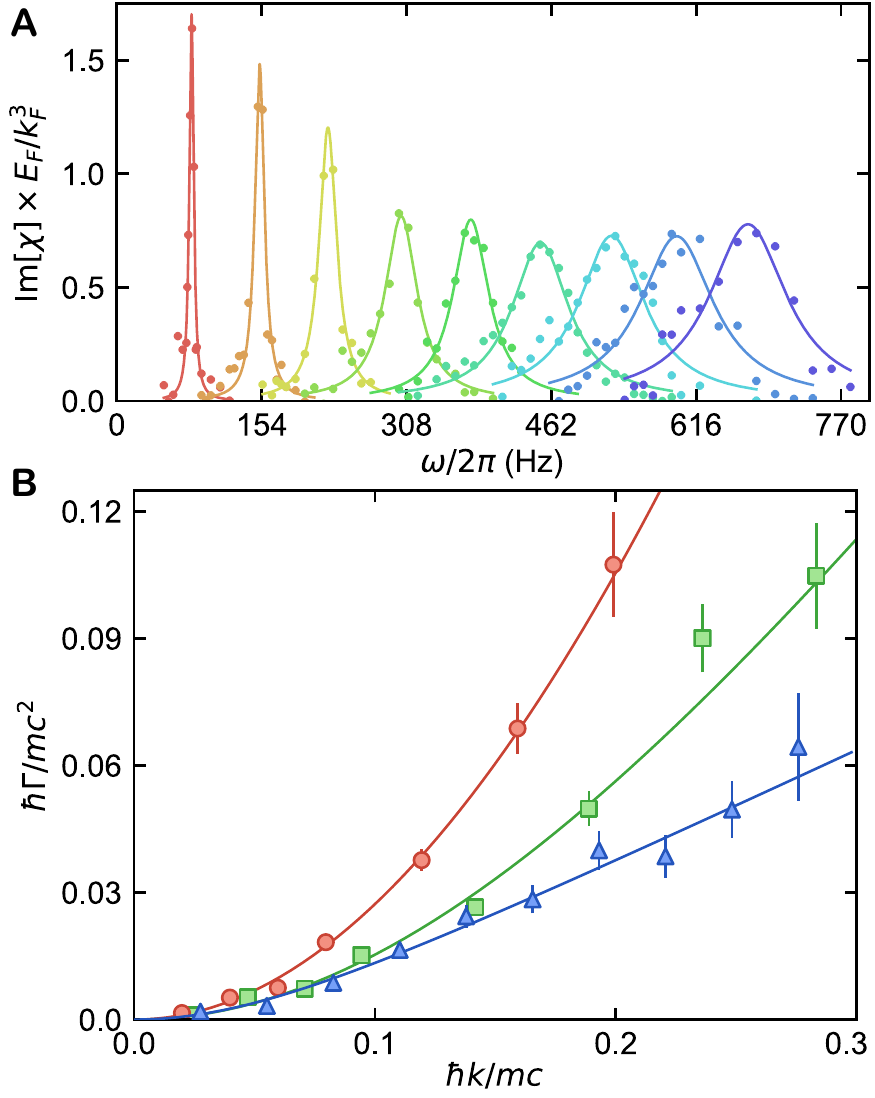}
\caption{\textbf{Spectral response of sound and its attenuation rate.}
(\textbf{A}) The imaginary part of the density response function at each normal mode wavenumber $k_j$ displays a well-defined peak in frequency, whose full-width-at-half-maximum yields the mode damping rate $\Gamma$. This is obtained from a Lorentzian fit, shown by solid lines. (\textbf{B}) Damping rate $\Gamma(k)$ for gas temperatures $T/T_\textrm{F} =$ 0.36(5) (red circles), 0.21(3) (green squares) and 0.13(2) (blue triangles). 
For all temperatures, $\Gamma(k)$ displays the characteristic quadratic scaling at low momenta implied by diffusive damping. For our coldest samples, as $k$ increases we observe a deviation from this behaviour, revealed by a crossover to linear scaling. At all temperatures and wavenumbers, our data are well-captured by the model of~\cite{Pethick1966:SoundAttenHe} (solid lines) which accounts for the finite relaxation rate of the fluid.
\label{Fig3}}
\end{figure}

\begin{figure}
\includegraphics[width=\columnwidth]{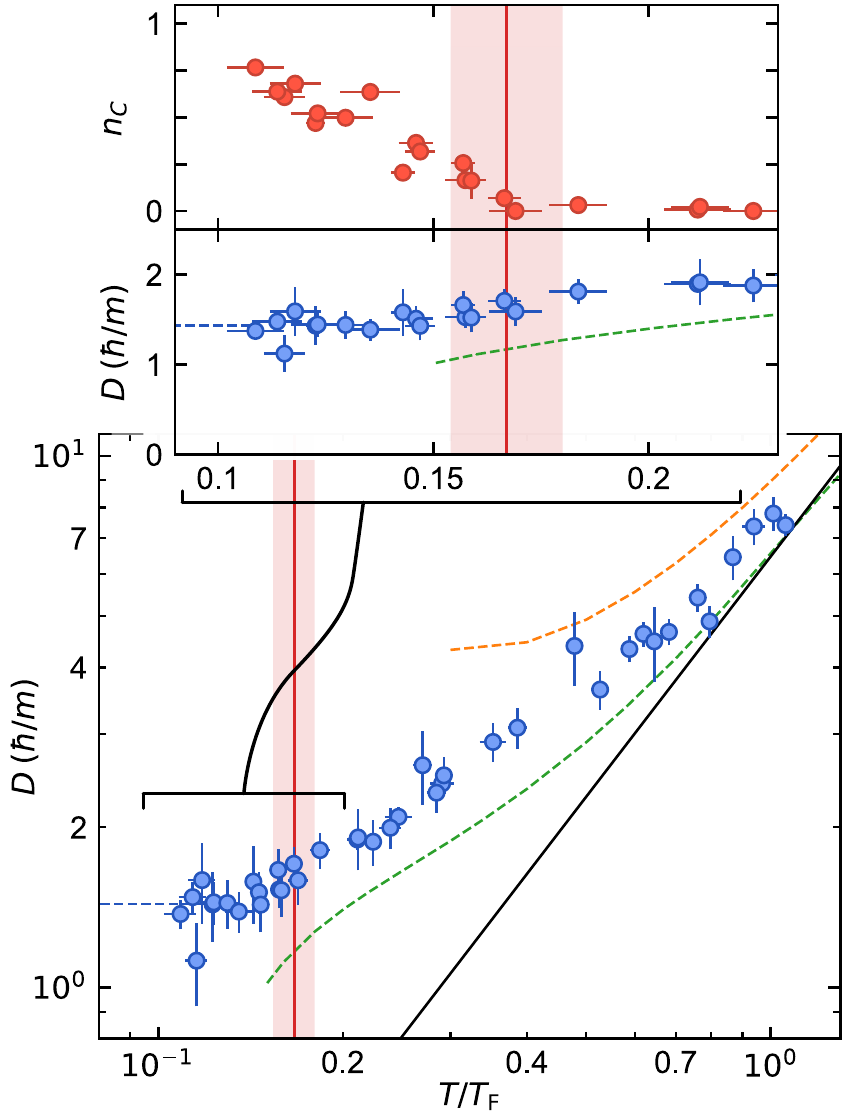}
\caption{\textbf{Temperature dependence of the sound diffusivity.} For temperatures comparable to the Fermi temperature, the sound diffusivity ($D$, normalized by $\hbar / m$; blue circles) approaches the expected high temperature scaling of $T^{3/2}$ (solid black line). As the temperature is lowered, $D$ decreases monotonically and attains a quantum-limited value close to $\hbar/m$. Below the superfluid transition (vertical red line, from~\cite{KuZwierlein2012:EoS}), $D$ is observed to be almost independent of temperature and condensate fraction ($n_C$, red circles). From the transition temperature ($n_C = 0$) to the coldest temperatures ($n_C \sim 0.8$), the changes in $D$ are within the standard error of the measurements. Theoretical predictions for $D$: the dashed orange line is from the sound attenuation length calculated in the framework of kinetic theory~\cite{BrabySchafer2010:ThermalCondSoundAtten} and the dashed green line is from a calculation of shear viscosity~\cite{EnssZwerger2011:Viscosity} assuming a Prandtl number of $2/3$.
}
\end{figure}

The density response $\mathrm{Im} [\chi(\omega, k_j)]$ at the wavenumber ${k_j=j \pi/L}$ of the $j^{\mathrm{th}}$ normal mode of the box is shown in Fig.~3A, along with Lorentzian fits (solid lines). The full-width-at-half-maximum yields the damping rate of sound $\Gamma$, which is shown as a function of $k$ in Fig. 3B, for gases both above (red and green) and below (blue) the superfluid transition. 
At temperatures above the transition to superfluidity, ${T> T_{\rm C} = 0.17~T_\textrm{F}}$~\cite{KuZwierlein2012:EoS} we observe $\Gamma(k)$ to increase quadratically with $k$ for all explored wave numbers ($k\lesssim 0.3\,mc/\hbar$). This establishes diffusive damping of sound in the normal regime, as expected in the collisionally hydrodynamic regime~\cite{Thomas2002:HydroAnisotropicExpansion,WrightGrimm2007:CollectiveFermiCrossover}.

Below the superfluid transition temperature, ${T<T_{\rm C}}$, we observe a crossover from quadratic scaling of $\Gamma(k)$ at wave numbers ${k\lesssim 0.2 mc/\hbar}$ to linear behaviour, indicating a departure from purely hydrodynamic transport at high wave numbers.
This is expected when the modulation frequency becomes comparable to the damping rate of thermal phonons $\Gamma_{\textrm{ph}}$~\cite{Pethick1966:SoundAttenHe,KurkjianCastin2017:PhononInteractions}. 
Collisionless or Landau damping of sound is due to non-linearities resulting from the kinetic energy density carried by sound and the density dependence of the speed of sound. Fermi's Golden Rule yields a rate $\Gamma_{\textrm{ph}} \propto k$~\cite{Hohenberg1965:SFHeTheory,KurkjianCastin2017:PhononInteractions} proportional to the energy $\hbar c k$ carried by a phonon. Including a non-zero damping rate of phonons $\Gamma_{\textrm{ph}}$ yields a crossover from hydrodynamic to collisionless damping as the sound frequency $c k$ exceeds $\Gamma_{\textrm{ph}}$~\cite{Pethick1966:SoundAttenHe}.
The relation $\Gamma = Dk^2 f(c k / \Gamma_{\mathrm{ph}})$ with $f(x) = \tan^{-1}(x)/x$~\cite{Pethick1966:SoundAttenHe} shows a good agreement with the data (solid line), with $\Gamma_{\textrm{ph}} =0.27(8)~k_B T / \hbar$ hinting at quantum critical damping~\cite{Enss2012:QmCritTransportUFG}.
We note that the observation of quadratic scaling of $\Gamma$ with $k$ at low wave numbers implies that sound is primarily attenuated in the bulk, and that edge effects are negligible~\cite{Wolfle1980:Viscosity3He,LL:FluidMechanics}.


As the main result of this work, we present in Fig.~4 the sound diffusivity ${D=\Gamma/k^2}$ of the unitary Fermi gas, obtained from the damping of low momentum sound modes. The measured values are expressed in units of $\hbar/m$, demonstrating universal sound diffusion.

Generally, the sound diffusivity contains contributions from both the bulk and shear viscosity, $\zeta$ and $\eta$ respectively, (which damps momentum gradients), and the thermal conductivity $\kappa$ (which damps temperature gradients)~\cite{LL:FluidMechanics}.
However, for a scale invariant fluid, the bulk viscosity vanishes~\cite{Son2007:VanishBulkVisc} and $D = D_\eta + D_\kappa$ only, with $D_\eta = 4 \eta/ (3 m n)$ and $D_\kappa =  4 \kappa T/(15 P)$ (see Supplementary Information).
We note that our measurements of $D$ therefore constrain the relationship between the viscosity and thermal conductivity, which is usually quantified by the Prandtl number $\mathrm{Pr} = c_P \eta / \kappa$~\cite{LL:FluidMechanics}, where $c_P$ is the specific heat at constant pressure (see Supplementary Information). 

The solid black line in Fig.~4 shows a prediction $D=6.46\,(\hbar/m)(T/T_\textrm{F})^{3/2}$, which uses the high-temperature results for viscosity~\cite{Bruun2005:Viscosity,EnssZwerger2011:Viscosity} and thermal conductivity~\cite{EnssZwerger2011:Viscosity,BrabySchafer2010:ThermalCondSoundAtten}, along with the ideal gas equation of state. 
This simple model captures the high-temperature behaviour well without any free parameters.
However, it is expected to underestimate $D$ when $T/T_\textrm{F}\lesssim 1$ since it neglects the suppression of scattering arising from Pauli blocking.

As the temperature is reduced, $D$ smoothly drops to a value ${\sim \hbar/m}$, consistent with Heisenberg-limited diffusivity.
Notably, at intermediate temperatures, we neither observe the ${D\sim 1/T^2}$ scaling typical of a Fermi liquid, nor any sudden change at the superfluid transition. 
This is further demonstrated by the inset of Fig.~4, where we show a magnified plot of $D$ (blue points) in the vicinity of the superfluid transition (vertical red line)~\cite{KuZwierlein2012:EoS}. 
Also shown is the measured pair condensate fraction (red points), which both indicates superfluidity and provides a robust thermometer in the superfluid phase~\cite{Zwierlein2014:NovelSuperfluids}. 
Despite the definitive onset of pair condensation, we observe no measurable sharp feature in the diffusivity, which remains approximately constant as the temperature is reduced. 

This behaviour can qualitatively be understood as follows. 
In the superfluid phase, viscosity arises entirely from
the normal component, i.e. the gas of thermal excitations, giving a diffusivity ${D\sim (n_{\rm n}/n) l v }$, where $n_{\rm n}$ is the density of the normal component, $l$ the mean free path of thermal excitations, and $v$ their average velocity~\cite{Pethick1977,TransportPhenomenaBook:JensenSmith}. 
At the temperatures studied here, the normal component is dominated by broken pairs~\cite{Zwierlein2014:NovelSuperfluids} of number density ${n_{\rm ex} \propto e^{-\Delta/k_\textrm{B} T}}$, with ${\Delta \approx 0.4~E_F}$ the pairing gap~\cite{SchirotzekKettrle2008:GapRF}, and thus ${n_{\rm n} \propto n_{\rm ex}}$ is exponentially low. However, the mean-free path between collisions is at the same time exponentially large, so that $D$ is independent of $n_{\rm ex}$ and close to its value at ${T=T_c}$~\cite{Pethick1977,TransportPhenomenaBook:JensenSmith}. Given a Fermi distribution broadened by $\Delta$ around the Fermi momentum, Pauli blocking will result in a value of ${D \sim  \frac{\hbar}{m}\frac{E_F^2}{\Delta^2}}$. 
In the unitary Fermi gas, ${\Delta\sim  E_\textrm{F}}$~\cite{SchirotzekKettrle2008:GapRF,HoinkaVale2017:GoldstonePairBreakingExcitationInUFG}, giving a diffusivity ${D\sim\hbar/m}$, consistent with our observations.
In contrast, the pairing gap in $^3$He is $\Delta/E_F{\sim \! 10^{-3} E_\textrm{F}}$, leading to a much larger value of ${D \sim 5,\!000~\hbar/m}$~\cite{Ono1982:3HeViscosity,Wolfle1980:Viscosity3He}.

In conclusion, we have measured the sound diffusivity of the unitary Fermi gas, with direct relevance to the flow of neutron matter at 25 orders of magnitude higher density.
The diffusivity approaches a Heisenberg-limited value of $\hbar/m$ at low temperatures, similar to the strongly interacting, bosonic quantum fluid $^4$He.
In contrast to Fermi liquid behavior seen in weakly interacting fermionic systems, the diffusivity monotonically increases with increasing temperatures and eventually follows the high-temperature behavior ${D\sim \hbar/m (T/T_F)^{3/2}}$.
The measured sound diffusivity constrains the shear viscosity and thermal conductivity of the unitary Fermi gas. In particular, combined with the calculated kinematic viscosity in~\cite{EnssZwerger2011:Viscosity} we find a Prandtl number strictly lower than unity for all explored temperatures (see Supplementary Information). 
This excludes the existence of a relativistic conformal gravity dual of the unitary Fermi gas~\cite{RangamaniSon2009:ConformalGravityDual}, as this would require $\mathrm{Pr} = 1$.
Thanks to the scale invariance of the unitary Fermi gas, the results obtained here apply broadly to other strongly interacting forms of fermionic matter, from hydrodynamic electron flow to nuclei and neutron matter.

We thank Y. Castin, T. Enss, T. Sch{\"a}fer, C. J. Vale, and W. Zwerger for helpful discussions. 
This work was supported by the National Science Foundation (Center for Ultracold Atoms Awards No. PHY-1734011 and No. PHY- 1506019), Air Force Office of Scientific Research (FA9550-16-1-0324 and MURI Quantum Phases of Matter FA9550-14-1-0035), Office of Naval Research (N00014-17-1-2257) and the David and Lucile Packard Foundation. J. S. was supported by LabEX ENS-ICFP: ANR-10-LABX-0010/ANR-10-IDEX-0001-02 PSL*.

\bibliography{main}

\onecolumngrid
\pagebreak

\part{}
\title{Title }
\maketitle

\section*{U\lowercase{niversal} S\lowercase{ound} D\lowercase{iffusion in a} S\lowercase{trongly} I\lowercase{nteracting} F\lowercase{ermi} G\lowercase{as} \protect\\ S\lowercase{upplementary} I\lowercase{nformation}}

\noindent
\textbf{Sample preparation and sound injection.}
The strongly interacting unitary Fermi gas was realized using an equal mixture of the first and third lowest hyperfine states of $^6$Li, ${\ket{1} = \ket{m_J = -\frac{1}{2}, m_I=1}}$ and ${\ket{3} = \ket{-\frac{1}{2}, -1}}$ respectively, with magnetic fields tuned to an interstate Feshbach resonance centered at ${\sim \! 690}$ G~\cite{Chin2010:FBResonance,ZurnHutson2013:PreciseFeshbachRes6Li}. The temperature and density were calibrated using the measured equation of state~\cite{KuZwierlein2012:EoS}. 
Sound waves were generated by sinusoidally modulating the intensity of one of the endcap laser sheets with sharpness ${\sim 4~\mu}$m~\cite{MukherjeeZwierlein2017:HomogFermi}. This drives the gas at a wide range of wavenumbers $({k \lesssim   0.5~\mu \mathrm{m}^{-1}} \textrm{~or~} {k/k_\textrm{F} \lesssim   0.15})$ simultaneously. The sound wave amplitude $\Delta n / n$ was deliberately kept below $10 \%$ to ensure that the response is in the linear regime and the local velocity ${v = (\Delta n / n)~c}$ is smaller than the critical velocity.
\\

\noindent
\textbf{Thermal conductivity and Prandtl number.}
Within hydrodynamics, the change in the energy of a sound wave is given by ${\Dot{E} = - D k^2 E}$ with $D = {4 \eta} /  ({3 \rho}) + {\alpha^2 c^2 \kappa T}/ ({ \rho c_P^2}) $~\cite{LL:FluidMechanics}. Here ${\left. \alpha = (1/V) (\partial V/\partial T) \right|_P}$ is the thermal expansivity and ${\rho = m n}$ is the mass density. The scale invariance of the unitary Fermi gas implies ${c^2 = 5P/ (3\rho)}$ and ${c_P = 5 P \alpha / (2 \rho)}$~\cite{Zwierlein2014:NovelSuperfluids}, which simplifies the sound diffusivity to $D = {4 \eta} /  ({3 \rho}) + 4 \kappa T / (15 P)$, valid at all temperatures above $T_c$. Below $T_c$, coupling to the second sound increases the contribution from viscosity by ${\sim 30\%}$ for the unitary Fermi gas \cite{Hohenberg1965:SFHeTheory}.

Our measurements of the sound diffusivity constrain the value of the viscosity and thermal conductivity according to $D = D_\eta + D_\kappa$, where $D_\eta = 4\eta / (3 \rho)$ and $D_\kappa = 4 \kappa T / (15 P)$.
We calculate the thermal conductivity $\kappa$ (Fig.~5A) and Prandtl number $\mathrm{Pr} = c_P \eta / \kappa$ (Fig.~5B) using the measured sound diffusivity, the experimental equation of state~\cite{KuZwierlein2012:EoS}, and a theoretical calculation for the shear viscosity $\eta$ above $T_c$~\cite{EnssZwerger2011:Viscosity}, performed within the same framework that gave excellent agreement with the experimental equation of state~\cite{KuZwierlein2012:EoS}. Similar to the sound diffusivity and viscosity, the thermal conductivity increases with temperature as $T^{3/2}$ for $T \gg T_\textrm{F}$.
The solid black line in Fig.~5A shows the limiting behavior ${ \kappa / (n k_\textrm{B}) = 10.38 \left( \hbar/m \right) \left( T/T_\textrm{F} \right)^{3/2} }$ for the thermal conductivity at high temperatures~\cite{BrabySchafer2010:ThermalCondSoundAtten}, which captures our data well without any free parameters. 

The Prandtl number, $\mathrm{Pr}$, quantifies the relative importance of viscosity and thermal conductivity for the attenuation of sound in fluids. 
For compressible fluids such as air, both the viscosity and the thermal conductivity play an important role in the diffusion of sound, resulting in $\mathrm{Pr}$ being close to unity. 
In contrast, for incompressible fluids such as water, thermal gradients associated with sound waves are minimal, resulting in a $\mathrm{Pr} \gg 1$. 
The unitary Fermi gas is a compressible fluid, whose $\mathrm{Pr}$ is predicted to reach the classical limit of $2/3$ at high temperatures~\cite{BrabySchafer2010:ThermalCondSoundAtten}. Our data indeed approach this value at high temperatures. We find the $\mathrm{Pr}$ to be significantly below $1$ at all temperatures, excluding the existence of a relativistic conformal gravity dual of the unitary Fermi gas~\cite{RangamaniSon2009:ConformalGravityDual}. 

\begin{figure}[H]
\includegraphics[width=1\columnwidth]{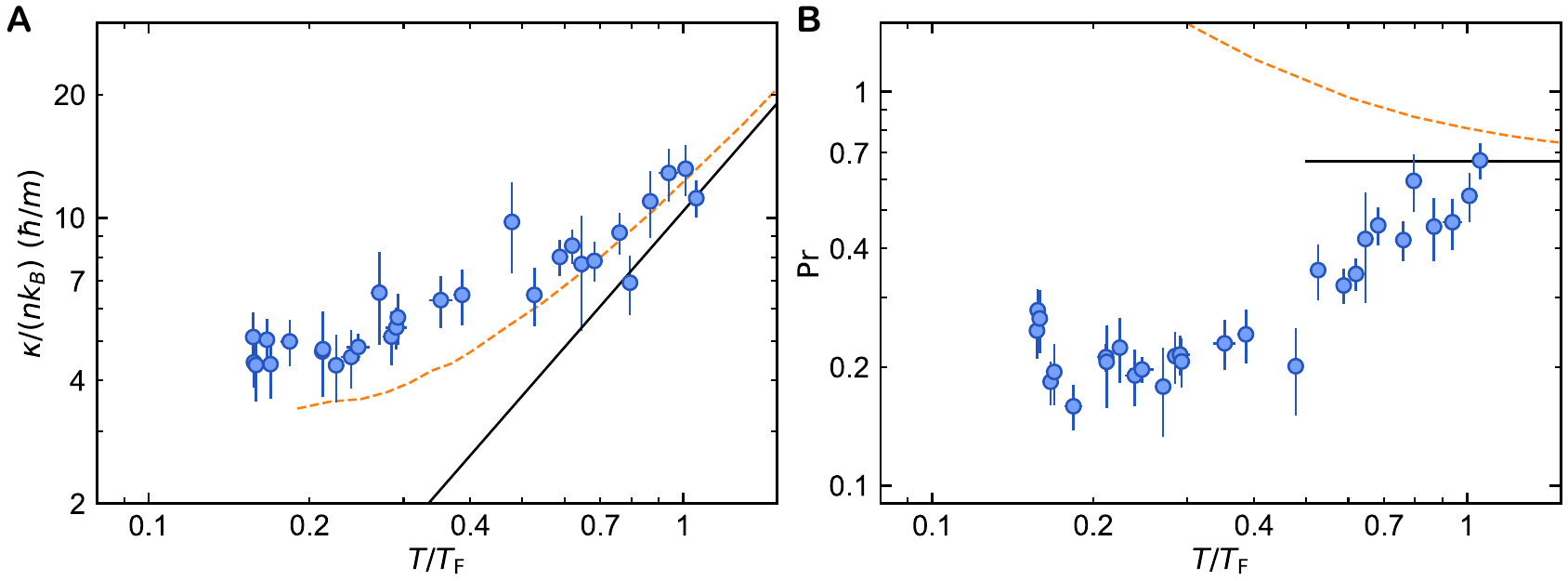}
\caption{\textbf{Thermal conductivity and Prandtl number.} For temperatures comparable to the Fermi temperature $({T \sim T_\textrm{F}})$, the thermal conductivity (\textbf{A}; $\kappa / (n k_\textrm{B})$, normalized by $\hbar / m$; blue circles) approaches the expected high temperature scaling $T^{3/2}$ (solid black line) as the Prandtl number (\textbf{B}; $\mathrm{Pr}$; blue circles) approaches the predicted high temperature value of $2/3$ (solid black line). The orange dashed line in both \textbf{A} and \textbf{B} are theoretical predictions calculated in the framework of kinetic theory~\cite{BrabySchafer2010:ThermalCondSoundAtten}. 
\label{Fig5}}
\end{figure}

\noindent
\textbf{The response function $\boldsymbol{\chi}$ and its normalization.}
The response function $\chi$ relates the perturbations in a fluid's number density to the applied external potential, $\delta n (\omega, k) = \chi(\omega, k) V(\omega, k)$~\cite{Hohenberg1965:SFHeTheory}. 
With knowledge of $\chi$, the density response of a fluid to an arbitrary external perturbation can be calculated via a Fourier transform, $ \delta n (t, x) = \int \frac{d \omega^\prime }{ 2 \pi} e^{- i \omega^\prime t} \int \frac{ d k }{ 2 \pi} e^{- i k x} ~ \chi(\omega^\prime, k) V(\omega^\prime, k) $. 
For example, the density response to a sinusoidal drive $V(\omega^\prime, k) = - i \pi V_0(k) \left(\delta(\omega^\prime + \omega) - \delta(\omega^\prime - \omega) \right)$, with frequency $\omega$ and amplitude $V_0(k)$, is 
\begin{equation*}
    \delta n (t, k) 
    = V_0 (k) \sin(\omega t) \mathrm{Re} [\chi(\omega, k)]  \allowbreak - V_0 (k) \cos(\omega t) \mathrm{Im} [\chi(\omega, k)] . 
\end{equation*}
Similar to a classical harmonic oscillator, the in-phase and out-of-phase density responses are proportional to $\mathrm{Re}[\chi]$ and $\mathrm{Im}[\chi]$ respectively, providing an experimentally convenient tool to measure the density response function. 
Data shown in Fig.~2 were taken after 30 complete cycles of the $\sin(\omega t)$ drive, which was found to be sufficiently long to reach a steady state at all frequencies and temperatures studied. 

In the vicinity of a sound mode $(\omega \sim c k)$, the response of the fluid can be well modeled by a damped driven harmonic oscillator with a resonance frequency $\omega_0 = c k$ and damping rate $\Gamma$~\cite{Hohenberg1965:SFHeTheory}. 
The equation of motion of a harmonic oscillator implies a response function $\chi \sim 1 / \left( \omega^2 - \omega_0^2 + i \Gamma \omega \right)$ whose imaginary part, $\mathrm{Im} [\chi] \sim 1 / ( (\omega - \omega_0)^2 + \Gamma^2 )$, has a Lorentzian peak with full-width-at-half-maximum $\Gamma$ centered at $\omega = \omega_0$. 

In general, the response function $\chi$ for a unitary Fermi gas is given by two-fluid hydrodynamics, as  discussed in Refs.~\cite{Hohenberg1965:SFHeTheory,Hohenberg1973:DensityCorrFunc,Hu2018:SoundChiS}. 
Fixing $k$, the function $\mathrm{Im}(\chi(\omega,k))/\omega$ contains in general two peaks: one is centered at the first-sound resonance, corresponding to predominantly density waves. The second peak is present when thermal gradients can cause density gradients, which occurs for non-zero expansivity $\alpha$ or equivalently for a specific heat ratio $c_P/c_V \ne 1$. In the normal state, the second peak occurs at zero frequency, corresponding to purely diffusive heat transport coupled to density. In the superfluid regime, this peak moves to finite frequency, corresponding to the emergence of second sound. It is predominantly (for $c_P/c_V$ not far from 1) a temperature wave that propagates ballistically~\cite{Bertaina2010,Sidorenkov2013}.

An exact sum rule relates the integral of $\mathrm{Im}(\omega,k)/\omega$ to the isothermal compressibility~\cite{Hohenberg1965:SFHeTheory,Hu2018:SoundChiS}.
The integral $W_1 = \int d\omega~\mathrm{Im} [\chi(\omega, k)]/\omega$ over only the first-sound peak is $n\pi / (2 m c^2)$, related to the speed of sound and thus the isentropic compressibility, independent of the wavenumber. 
We verify this `first sound sum-rule' in Fig.~6 and utilize it to calibrate the amplitude $V_0(k)$ of the drive. 
The measured out-of-phase density response (Fig.~2C and Fig.~3A) is given by $\delta n (\omega, k) = \mathrm{Im} [\chi (\omega, k)] V_0(k)$. 
The weight of the first-sound mode is calculated from the density response, $W_1 = \left[ \int d\omega~\delta n (\omega, k) / \omega  \right]/V_0(k) \equiv I_1 / V_0(k)$, where $I_1$ is the integral over the $\delta n$ (Fig.~6A).
We model the shape of the potential wall $V_0(x)$ by a Gaussian function with a width $\sigma$ such that $V_0(k) \sim \exp{\left[ - k^2 \sigma^2 / 2 \right] }$. 
To account for the slight asymmetry between the two endcap potentials, we use $\sigma = 4.4(1) \mu$m and $3.2 \mu$m for the even and odd modes, respectively, acquired from Gaussian fits to $I_1$. 
The calculated weight $W_1(k)$ (Fig~6B) is independent of the wavenumber to within the standard error of the measurements.  
By requiring the average value of $W_1(k)$ to be $n \pi / (2 m c^2)$, we calibrate the amplitude of $V_0(k)$ and normalize $\mathrm{Im} [\chi]$ shown in Fig.~3A.  

\begin{figure}
\includegraphics[width=1\columnwidth]{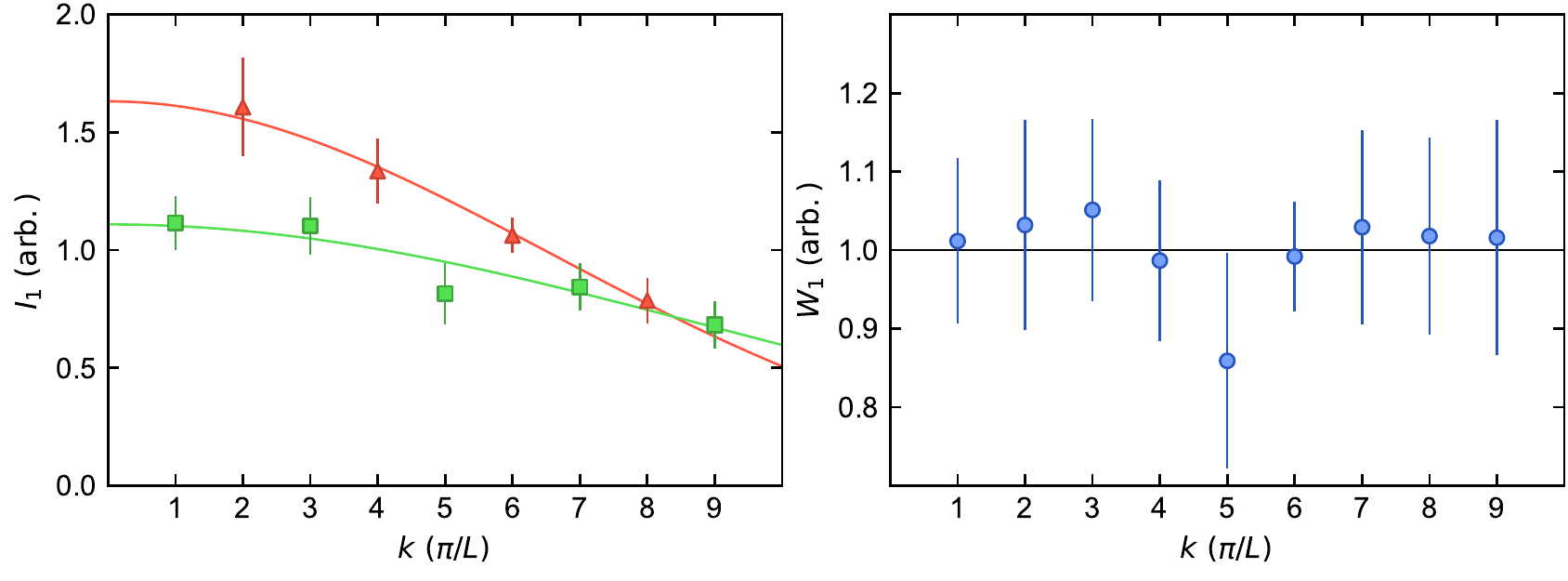} 
\caption{\textbf{Weight of the first-sound mode in $\boldsymbol{\chi}$. } 
(\textbf{A}) Integral over the density response $I_1(k) = \int d\omega~\delta n (\omega, k) / \omega$ for even (red triangles) and odd (green square) modes. They are fit with a Gaussian function (solid lines) which models the drive $V_0(k)$.
(\textbf{B}) The weight of the first sound in the density response function $W_1(k) = \int d\omega~\mathrm{Im} [\chi(\omega, k)]/\omega$ calculated from the measured $I_1$ and the modelled drive potential, $W_1(k) = I_1(k) / V_0(k)$. 
\label{Fig7}}
\end{figure}






\end{document}